# Measurement-to-track association and finite-set statistics

Ronald Mahler, Eagan MN, USA, EML: mahlerronald@comcast.net

## Abstract

This is a shortened, clarified, and mathematically more rigorous version of the original arXiv version. Its first four main findings remain unchanged from the original: 1) measurement-to-track associations (MTAs) in multitarget tracking (MTT) are heuristic and physically erroneous multitarget state models; 2) MTAs occur in the labeled random finite set (LRFS) approach only as purely mathematical abstractions that do not occur singly; 3) the labeled random finite set (LRFS) approach is not a mathematically obfuscated replication of multi-hypothesis tracking (MHT); and 4) the conventional interpretation of MHT is more consistent with classical than Bayesian statistics. This version goes byond the original in including the following additional main finding: 5) a generalized, RFS-like interpretation results in a correct Bayesian formulation of MHT, based on MTA likelihood functions and MTA Markov transitions.

## 1. Introduction

When Reid introduced the multiple-hypothesis tracking (MHT) approach to multiple-target tracking (MTT) in 1979 [8], computer processing was primitive by today's standards. MHT partially sidestepped this limitation by using measurement-to-track associations (MTAs) to decompose MTT problems into systems of extended Kalman filters. Since then, advances by Reid's successors have made MHTs the workhorses of MTT. Understandably, however, limited computing power, combined with a lack of suitable mathematical machinery, also impeded proper theoretical rigor. But now, given the existence of greatly improved computational and mathematical tools, it is important to provide MTT with secure theoretical foundations that provide the basis for superior practical MTTs.

Labeled random finite set (LRFS) theory provides both. The RFS IDs/labels model appeared in 1997 in [2, p. 135, 196-197] and subsequently in [7, Sec. 14.5.6]; was systematically expanded into LRFS theory in [13], [14]; which has since been widely adopted. The first general exact closed-form approximation[1] of any version of the multitarget Bayes recursive filter (MTBRF)—the GLMB filter—also appeared in [13], a surprising discovery that has been widely adopted or emulated. The first use of Gibbs sampling in LRFS (and probably in MTT) was introduced in [3] followed by [15] and is being increasingly adopted or emulated.

The latest Gibbs-based GLMB filter implementations can near-optimally and simultaneously track over a million 3D targets in real time in significant clutter using off-the-shelf computing equipment [1], another surprising development. Also, GLMB-type filters have: quantifiable approximation errors [14]; linear complexity in the number of measurements [14]; log-linear complexity in the number of hypothesized tracks [10]; and linear complexity in the number of scans in the multi-scan case [12].

This is a shortened, clarified, and mathematically more rigorous version of the original arXiv version. Its first four main findings remain unchanged from the original:

1. MTAs are heuristic and physically erroneous multitarget state models (Section 8).

2. MTAs occur in LRFS, but as purely mathematical abstractions that do not occur singly (Section 9).

3. LRFS is not, as sometimes suggested, a mathematically obfuscated replication of MHT (Section 10).

4. The conventional interpretation of MHT is more consistent with classical than Bayesian statistics (Section 5.1).

This version goes beyond the original in including the following additional main finding:

5. A generalized, RFS-like interpretation results in a correct Bayesian formulation of MHT, based on MTA likelihood functions and MTA Markov transitions (Sections 5.2, 5.3, 6, 7).

What follows are concise summaries of: Bayesian vs. classical statistics (Section 2); LRFS (Section 3); MHT and MTAs (Section 4); MHT conceptual difficulties (Section 5); likelihood functions for MTAs (Section 6); Markov transitions for MTAs (Section 7); the physical reality of MTAs (Section 8); MTAs in MHT vs. MTAs in LRFS (Section 9); and Conclusions (Section 10).

[1] In the sense of [6].

## 2. Bayesian vs. Classical Statistics

Suppose that physical entities with states $\xi \in \mathcal{X}$ are observed by a sensor that collects measurements $\zeta \in \mathcal{Z}$. The goal is to estimate $\xi$ given $\zeta$. Bayesian and classical statistics both employ a likelihood function (measurement distribution) $L_\zeta(\xi) = f(\zeta|\xi)$ that specifies the probability (or probability density) that measurement $\zeta$ will be collected from $\xi$. Thus $\int f(\zeta|\xi)d\zeta = 1$ for all $\xi$.

2.1 *Bayesian Statistics.* The unknown state $\xi$ is presumed to be a random variable (RV) whose statistics are characterized by a prior distribution, $f_0(\xi)$. An estimate of $\xi$ is determined using some Bayes-optimal state estimator,[2] for example the maximum a posteriori (MAP) estimator:[3]

$$\hat{\xi} = \arg\sup_{\xi \in \mathcal{X}} f(\xi|\zeta) \tag{1}$$

where the posterior distribution on $\xi$ is $f(\xi|\zeta) = f(\zeta|\xi)\, f_0(\xi)/f(\xi)$ where $f(\xi) = \int f(\zeta|\xi)\, f_0(\xi) d\xi$.

2.2 *Classical Statistics.* The unknown state $\xi$ is presumed to be a deterministic constant. From a Bayesian perspective, its distribution is $f_0(\xi) = \delta_{\xi_0}(\xi)$ where $\delta_{\xi_0}(\xi)$ is the Dirac delta concentrated at the unknown constant $\xi_0$. An estimate $\hat{\xi}$ of $\xi$ is determined using the maximum likelihood estimator (MLE):

$$\hat{\xi} = \arg\sup_{\xi \in \mathcal{X}} f(\zeta|\xi). \tag{2}$$

## 3. LRFS

In LRFS $\mathcal{X} = \mathcal{F}_{\mathbb{L}}(\mathbb{X}_0)$ where $\mathcal{F}_{\mathbb{L}}(\mathbb{X}_0)$ is the set of all *labeled finite subsets* (LFSs)

$$X = \{(x_1, l_1), ..., (x_n, l_n)\} \tag{3}$$

of $\mathbb{X} = \mathbb{X}_0 \times \mathbb{L}$, where $x_1, ..., x_n \in \mathbb{X}_0 = \mathbb{R}^N$ are the targets' kinematic states and the *distinct* labels $l_1, ..., l_n \in \mathbb{L}$ *uniquely identify* them. Let $0 \le |S| \le \infty$ denote the number of elements in a set $S$ and $X_{\mathbb{L}} = \{l_1, ..., l_n\}$ denote the set of labels in $X$. Then $X$ is LFS if and only if $|X| = |X_{\mathbb{L}}|$. A label is a symbol for a discrete state variable: target identity. If there are no labels (or, equivalently, if $|\mathbb{L}| = 1$) then $\mathcal{F}_{\mathbb{L}}(\mathbb{X}_0) = \mathcal{F}(\mathbb{X}_0)$. A labeled random finite subset (LRFS) $\Xi \subseteq \mathbb{X}$ is an RV on $\mathcal{F}_{\mathbb{L}}(\mathbb{X}_0)$.

Now consider LRFS with the "standard multitarget measurement model" (SMMM), in which a detection process is applied to to a sensor signature [7, Sec. 12.3], [4, Sec. 7.2]. Then $\mathcal{Z} = \mathcal{F}(\mathbb{Z}_0)$ is the set of all finite subsets $Z$ of the single-target, single-sensor measurement space $\mathbb{Z}_0$. The likelihood has the form $L_Z(X) = f(Z|X)$, where $\int f(Z|X)\delta Z = 1$ for all LFSs $X$ and $\int \cdot \delta Z$ is an RFS set integral:

$$\int f(Z)\delta Z = \sum_{m \ge 0} \frac{1}{m!} \int f(\{z_1, ..., z_m\}) dz_1 \cdots dz_m. \tag{4}$$

The MAP estimator does not exist and must be replaced by (for example) the Bayes-optimal "JoM" estimator [7, Sec. 14.5], [4, Sec. 5.3]

$$\hat{X} = \arg\sup_X \frac{c^{|X|} f(X|Z)}{|X|!}, \tag{5}$$

where $c > 0$ is a small constant with the same unit of measurement as $\mathbb{X}_0$; and where the posterior distribution is $f(X|Z) = f(Z|X)\, f_0(X)/f(Z)$ with $f(Z) = \int f(Z|X)\, f_0(X)\delta X$.

## 4. MHT and MTAs

MHT is vector-based rather than set-based, and presumes the SMMM. Its multitarget states are MTAs rather than LFSs.

In this case $\mathcal{X} = \vec{\mathbb{X}}_0 = \uplus_{n\ge0}\mathbb{X}_0^n$ where $\mathbb{X}_0^n$ is the $n$'th Cartesian power of $\mathbb{X}_0$ and "$\uplus$" indicates disjoint union; and where $\mathbb{X}_0^0 = \{\phi\}$ where $\phi$ is the null (empty) vector. Likewise, $\mathcal{Z} = \vec{\mathbb{Z}}_0 = \uplus_{m\ge0}\mathbb{Z}_0^m$. The sets $X \in \mathcal{F}(\mathbb{X}_0)$ and $Z \in \mathcal{F}(\mathbb{Z}_0)$ are replaced by vectors $\vec{X} = (x_1, ..., x_n) \in \vec{\mathbb{X}}_0$ and $\vec{Z} = (z_1, ..., z_m) \in \vec{\mathbb{Z}}_0$ with *presumed orderings* $x_1, ..., x_n$ and $z_1, ..., z_m$.

In what follows, $\{\vec{Z}\} = \{z_1, ..., z_m\}$ will denote the set of entries[4] in $\vec{Z}$ and $|\vec{Z}|$ the number of those entries. In practice, if $\vec{Z} = (z_1, ..., z_m)$ then the measurements $z_1, ..., z_m$ will always be distinct—i.e., $|\{\vec{Z}\}| = |\vec{Z}| = m$.

A real-valued function $f(Z)$ on $Z \in \mathcal{F}(\mathbb{Z}_0)$ induces a real-valued function $\vec{f}(\vec{Z})$ on $\vec{Z} \in \vec{\mathbb{Z}}_0$ as follows:

$$\vec{f}(\vec{Z}) = \begin{cases} \frac{1}{|\vec{Z}|!} f(\{\vec{Z}\}) & \text{if} \quad |\vec{Z}| = |\{\vec{Z}\}| \\ 0 & \text{if} \quad \text{otherwise} \end{cases}. \tag{6}$$

Because $\mathbb{Z}_0 = \mathbb{R}^M$ is continuously infinite, the RFS set integral $\int \cdot \delta Z$ on $\mathcal{F}(\mathbb{Z}_0)$ is equivalent to the following integral $\int \cdot d\vec{Z}$ on $\vec{\mathbb{Z}}_0$:

$$\int f(Z)\delta Z = \int \vec{f}(\vec{Z}) d\vec{Z} = \sum_{m\ge0} \int \vec{f}(z_1, ..., z_m) dz_1 \cdots dz_m \tag{7}$$

where we have abbreviated $\vec{f}(z_1, ..., z_m) = \vec{f}((z_1, ..., z_m))$ for all $(z_1, ..., z_m) \in \vec{\mathbb{Z}}_0$.

4.1 *MTAs.* For conceptual clarity, assume that $1, ..., n$ identify all and the only targets present at all times, with spatial distributions of the form $s_1(x), ..., s_n(x)$. Let $\vec{Z} = (z_1, ..., z_m) \in \vec{\mathbb{Z}}_0$ be the measurement-vector collected by the sensor. An MTA for $\vec{Z}$ is a function

$$\alpha^{\vec{Z}} : \{1, ..., n\} \to \{0, 1, ..., m\} \tag{8}$$

[2]This is the only meaning of the term "Bayes-optimal." An algorithm is not Bayes-optimal simply because it employs Bayes' rule in some fashion.

[3]The MAP estimator is often described as "Bayes-optimal." In the case of continuously infinite state spaces, however, it is actually approximately Bayes-optimal.

[4]This is an abuse of notation. Strictly speaking, $\{\vec{Z}\}$ denotes a set with a single element $\vec{Z}$. The meaning $\{\vec{Z}\} = \{z_1, ..., z_m\}$ will be clear from context.

such that $\alpha^{\vec{Z}}(i) = \alpha^{\vec{Z}}(i') > 0$ implies $i = i'$. The set of all MTAs for $\vec{Z}$ is denoted $\mathbb{A}^{\vec{Z}}$ and the number of such MTAs is $\bar{N}_m$. Note that because $\alpha^{\vec{Z}}$ and the pair $(\alpha^{\vec{Z}}, \vec{Z})$ contain exactly the same information, the former is an abbreviation of the latter. Also, whereas $\vec{Z}$ is a constant, $\alpha^{\vec{Z}} \in \mathbb{A}^{\vec{Z}}$ is an RV with $\bar{N}_m$ possible realizations.

*Notation, 1*: An abbreviated notation will be employed in what follows. If $\vec{Z}_k$ is a measurement-vector collected at time $t_k$ then $\alpha_k^{\vec{Z}_k}$ will be abbreviated as $\alpha_k \in \mathbb{A}^{\vec{Z}_k}$, with $\vec{Z}_k$ being implicitly understood.

*Notation, 2*: Let $\vec{Z}_{1:k} : \vec{Z}_1, ..., \vec{Z}_k$ be a sequence of measurement-vectors at times $t_1, ..., t_k$ with, by convention, $\vec{Z}_{1:0} = \phi$ (null vector). Then $\alpha_{1:k} : \alpha_1, ..., \alpha_k$ with $\alpha_i = \alpha_i^{\vec{Z}_i}$ denotes a corresponding sequence of MTAs with, by convention, $\alpha_{1|0} : \{1, ..., n\} \to \{0\}$ being the trivial MTA. For a given $\vec{Z}_{1:k}$, the set of all such sequences is $\mathbb{A}^{\vec{Z}_{1:k}}$.

4.2 *Physical Interpretation of MTAs.* If $\alpha_k(i) = 0$ then the predicted target $i$ was not detected at time $t_k$; whereas if $\alpha_k(i) > 0$ then the measurement $z_{\alpha_k(i)}$ in $\vec{Z}_k$ was collected from it. For any $\vec{W} \in \vec{\mathbb{Z}}_0$ and any $\alpha^{\vec{W}} \in \mathbb{A}^{\vec{W}}$, denote the vector of target-generated measurements in $\vec{W}$ by:

$$\vec{W}^{\alpha} = (z_{\alpha(i)} \in \{\vec{W}\} : \alpha(i) > 0) \tag{9}$$

where the ordering of the entries in $\vec{W}^{\alpha}$ preserves the ordering in $\vec{W}$. Then $\{\vec{W}\} - \{\vec{W}^{\alpha}\}$ is the set of measurements in $\vec{W}$ not generated by any target—i.e, clutter and false detection measurements. In particular, if $\vec{W} = \vec{Z}_k$ then

$$\vec{Z}_k^{\alpha_k} = (z_{\alpha_k(i)} \in \{\vec{Z}_k\} : \alpha_k(i) > 0) \tag{10}$$

denotes the vector of target-generated measurements in $\vec{Z}_k$.

4.3 *HOMHT.* There are numerous variants of MHT. For the sake of conceptual clarity, Reid's "hypothesis-oriented" (HO) version [8] will be presumed—but in the following special case: target number $n$ is known and constant and the probability of detection $p_{D,k}(x)$ is state-dependent and usually nonconstant.

Suppose that measurement-vectors $\vec{Z}_{1:k-1} : \vec{Z}_1, ..., \vec{Z}_{k-1}$ have previously been collected and that MTAs $\alpha_{1:k-1} : \alpha_1, ..., \alpha_{k-1}$ have been chosen. Then the spatial distributions at time $t_{k-1}$ are $s_i^{k-1}(x) = s_i^{k-1}(x|\alpha_{1:k-1}, \vec{Z}_{1:k-1})$ for $i = 1, ..., n$. Each of these is predicted to time $t_k$,

$$s_i^{k|k-1}(x) = \int f_{k|k-1}(x|x_{k-1})\, s_i^{k-1}(x_{k-1}) dx_{k-1}, \tag{11}$$

where $f_{k|k-1}(x|x_{k-1})$ is the single-target Markov density.

Now let $\vec{Z}_k = (z_1, ..., z_{m_k})$ be the new measurement-vector collected at time $t_k$ and choose an MTA $\alpha_k : \{1, ..., n\} \to \{0, 1, ..., m_k\}$. Then the predicted spatial distributions are measurement-updated as follows:

$$s_i^k(x) \propto \begin{cases} \left( \begin{array}{c} \int p_{D,k}(x)\ f_k(z_{\alpha_k(i)}|x) \\ \cdot s_i^{k|k-1}(x_{k-1}) dx_{k-1} \end{array} \right) & \text{if} \quad \alpha_k(i) > 0 \\ (1 - p_{D,k}(x))\, s_i^{k|k-1}(x) & \text{if} \quad \alpha_k(i) = 0 \end{cases} \tag{12}$$

where $f_k(z|x)$ is the single-target measurement density at time $t_k$; and where $s_i^k(x) = s_i^k(x|\alpha_{1:k}, \vec{Z}_{1:k})$.[5]

Given this, the goal of HOMHT (known as "deferred decision logic") is to, at an appropriate time $t_j$ (when track uncertainties can most opportunely be resolved), determine what MTA-sequence $\hat{\alpha}_{1:k} : \hat{\alpha}_1, ..., \hat{\alpha}_k$ most accurately matches measurements with predicted targets over the time period $t_1, ..., t_k$ (see Section 4.5).

4.4 *Reid's Recursion.* Let $p_{1:k}(\alpha_{1:k}|\vec{Z}_{1:k})$ be the probability of the MTA-sequence $\alpha_{1:k}$ given the measurement-sequence $\vec{Z}_{1:k}$. Then Reid's recursion formula for these probabilities is the Bayesian update equation [8, Eq. 8]:[6]

$$\begin{aligned} & p_{1:k}(\alpha_{1:k}|\vec{Z}_{1:k}) \\ = & \frac{\vec{f}_{k|k-1}(\alpha_k, \vec{Z}_k|\alpha_{1:k-1}, \vec{Z}_{1:k-1})\, p_{1:k-1}(\alpha_{1:k-1}|\vec{Z}_{1:k-1})}{\vec{f}_k(\vec{Z}_k|\vec{Z}_{1:k-1})} \end{aligned} \tag{13}$$

with Bayesian transition density

$$\begin{aligned} \vec{f}_{k|k-1}(\alpha_k, \vec{Z}_k|\alpha_{1:k-1}, \vec{Z}_{1:k-1}) &= \vec{f}_k(\vec{Z}_k|\alpha_{1:k}, \vec{Z}_{1:k-1}) \\ & \quad \cdot p_{k|k-1}(\alpha_k|\alpha_{1:k-1}, \vec{Z}_{1:k-1}). \end{aligned} \tag{14}$$

Now apply the independence assumptions

$$p_{k|k-1}(\alpha_k|\alpha_{1:k-1}, \vec{Z}_{1:k-1}) = p_{k|k-1}(\alpha_k|\alpha_{k-1}) \tag{15}$$

$$\vec{f}_k(\vec{Z}_k|\alpha_{1:k}, \vec{Z}_{1:k-1}) = \vec{f}_k(\vec{Z}_k|\alpha_k), \tag{16}$$

where $p_{k|k-1}(\alpha_k|\alpha_{k-1}, \vec{Z}_{k-1}) = p_{k|k-1}(\alpha_k|\alpha_{k-1})$ since $\alpha_{k-1} = \alpha_{k-1}^{\vec{Z}_{k-1}}$ contains the same information as $(\alpha_{k-1}, Z_{k-1})$. Then $p_{k|k-1}(\alpha_k|\alpha_{k-1})$ is, seemingly, a conventional Bayesian Markov transition [7, Eq. 3.56]; and $\vec{f}_k(\vec{Z}_k|\alpha_k)$ is, seemingly, a conventional Bayesian measurement density [7, Eq. 3.54].

When $k = 1$, Eqs. (13, 14) reduce, also seemingly, to Bayes' rule:

$$p_1(\alpha_1|\vec{Z}_1) = \frac{\vec{f}_1(\vec{Z}_1|\alpha_1)\, p_{1|0}(\alpha_1)}{\vec{f}_1(\vec{Z}_1)}. \tag{17}$$

Here, $p_{1|0}(\alpha_1)$ is the Markov transition from the trivial MTA $\{1, ..., n\} \to \{0\}$ to $\alpha_1$—but which also serves as the initial distribution.

4.5 *Multitarget State Estimation.* At time $t_k$, the targets' spatial distributions will be $s_i^k(x|\alpha_{1:k}, \vec{Z}_{1:k})$ for $i = 1, ..., n$. From a practical point of view, these are meaningless unless the MTAs $\alpha_j \in \mathbb{A}^{\vec{Z}_j}$ for $j = 1, ..., k$ have been selected so as to best match measurements with targets—i.e., to resemble the "actual MTAs" as much as possible.[7] This selection is the MAP estimate:

$$\hat{\alpha}_{1:k} = \arg\max_{\alpha_{1:k}} p_k(\alpha_{1:k}|\vec{Z}_{1:k}). \tag{18}$$

[5] Note that if $p_{D,k}(x) = p_{D,k}$ is constant and $\alpha_k(i) = 0$ then $s_i^k(x) = s_i^{k|k-1}(x)$.

[6] In what follows, probability densities vs. probabilities will be denoted as $f(x)$ vs. $p(x)$ unless otherwise indicated. The former has a unit of measurement whereas the latter is unitless.

[7] In Section 8 it will be argued that "actual MTAs" are heuristic fictions.

where the $\arg\max$ is taken over the finite number of all $\alpha_{1:k} \in \mathbb{A}^{\vec{Z}_{1:k}}$. The multitarget state estimate at time $t_k$ is then $\hat{x}_i^k = \arg\sup_{x \in \mathbb{X}_0} s_i^k(x|\hat{\alpha}_{1:k}, \vec{Z}_{1:k})$ for $i = 1, ..., n$.

## 5. MHT Conceptual Difficulties

The transitions $s_i^{k-1}(x) \to s_i^{k|k-1}(x)$ and $s_i^{k|k-1}(x) \to s_i^k(x)$ for $i = 1, ..., n$ implicitly presume that the $n$ targets are sufficiently separated to be approximately independent. This is equivalent to the following approximation[8]

$$\vec{f}_k(x_1...., x_n|\alpha_{1:k}, \vec{Z}_{1:k}) \cong s_1^k(x_1|\alpha_{1:k}, \vec{Z}_{1:k}) \cdots s_n^k(x_n|\alpha_{1:k}, \vec{Z}_{1:k}). \tag{19}$$

But this also presumes a questionable interpretation of MTAs, as follows.

5.1 *The Conventional MHT Interpretation.* Implicit in Eq. (17) is the following assumption: $\vec{Z}_1$ and $m_1$ are known, and the randomness of $\alpha_1$, as characterized by $p_{1|0}(\alpha_1)$, involves only the variability of the MTA $\alpha_1 \in \mathbb{A}^{\vec{Z}_1}$. Thus the normalization condition for $p_{1|0}(\alpha_1)$ must be

$$\sum_{\alpha_1 \in \mathbb{A}^{\vec{Z}_1}} p_{1|0}(\alpha_1) = 1. \tag{20}$$

For the same reason, the MAP estimator must be $\hat{\alpha}_1 = \arg\max_{\alpha_1 \in \mathbb{A}^{\vec{Z}_1}} p_{1|0}(\alpha_1)$.

But neither can be true. If $p_{1|0}(\alpha_1)$ is an initial distribution then it encapsulates all information available *before* the collection $\vec{Z}_1 \sim \vec{f}_1(\vec{Z}|\alpha_1)$ of the first measurement-vector $\vec{Z}_1$. It thus cannot depend on $\vec{Z}_1$ or even $m_1$—which, in turn, chronologically contradicts the fact that $\alpha_1 = \alpha_1^{\vec{Z}_1}$.

Moreover, this interpretation is peculiar from a Bayesian perspective. If $\alpha_1^{\vec{Z}_1}$ is a multitarget state with state parameter $\vec{Z}_1$, then it, and $\vec{Z}_1$ in particular, are unknown RVs. But in actuality $\vec{Z}_1$ is known and constant—which means that the distribution of $\alpha^{\vec{Z}}$ must have the form $p_1(\alpha^{\vec{Z}}) = \delta_{\vec{Z}_1}(\vec{Z})\ p_1(\alpha^{\vec{Z}_1})$ where $\delta_{\vec{Z}_1}(\vec{Z})$ is the Dirac delta concentrated at $\vec{Z}_1$. This denouement is more consistent with classical statistics than Bayesian statistics (see Section 2).

5.2 *A General MHT Interpretation.* The Bayesian expressions $\vec{f}_1(\vec{Z}|\alpha^{\vec{Z}})$ and $p_{1|0}(\alpha^{\vec{Z}})$ imply that they are defined for *arbitrary* finite $\vec{Z} \in \vec{\mathbb{Z}}_0$ and *arbitrary* $\alpha^{\vec{Z}} \in \mathbb{A}^{\vec{Z}}$. That is, a probability $p_{1|0}(\alpha^{\vec{Z}})$ must be assigned to each $\vec{Z}$ and each $\alpha^{\vec{Z}} : \{1, .., n\} \to \{0, 1, ..., m\}$. Thus the true normalization condition must be[9]

$$\int \sum_{\alpha^{\vec{Z}} \in \mathbb{A}^{\vec{Z}}} p_{1|0}(\alpha^{\vec{Z}}) d\vec{Z} = 1, \tag{21}$$

where the RFS-like integral $\int \cdot d\vec{Z}$ is as defined in Eq. (7).

With this interpretation, the chronological contradiction disappears because the RV $\vec{Z}$ and thus $\vec{f}_1(\vec{Z}|\alpha_1)$ are no longer restricted to the realization $\vec{Z} = \vec{Z}_1$. This interpretation is what will be adopted hereafter. Its immediate consequence is that we must devise concrete definitions of $\vec{f}_k(\vec{Z}|\alpha^{\vec{Z}})$ and $p_{k|k-1}(\alpha^{\vec{Z}}|\alpha_{k-1})$ in Eqs. (15, 16) for all $\vec{Z} \in \vec{\mathbb{Z}}_0$ and, given $\vec{Z}$, for all $\alpha^{\vec{Z}} \in \mathbb{A}^{\vec{Z}}$. These definitions are the subjects of, respectively, Sections 6 and 7.

5.3 *State Estimation in the General Interpretation.* The rigorous version of Eq. (18) is the MAP estimate of $f_k(\alpha_{1:k}^{\vec{Y}_{1:k}}|\vec{Z}_{1:k})$ when $\vec{Y}_{1:k} = \vec{Z}_{1:k}$ is held constant:

$$\hat{\alpha}_{1:k}^{\vec{Z}_{1:k}} = \arg\sup_{\alpha_{1:k} \in A^{\vec{Z}_{1:k}}} \left[ p_k(\alpha_{1:k}^{\vec{Y}_{1:k}}|\vec{Z}_{1:k}) \right]_{\vec{Y}_{1:k} = \vec{Z}_{1:k}}. \tag{22}$$

## 6. Likelihood Functions for MTAs

6.1 *Extended MTAs for Likelihoods.* Define the extension of $\alpha_k = \alpha_k^{\vec{Z}_k} : \{1, ..., n\} \to \{0, 1, ..., m_k\}$ to $\alpha^{\vec{Z}} : \{1, ..., n\} \to \{0, 1, ..., m\}$ as follows:

$$\alpha(i) = \alpha^{\vec{Z}}(i) = \begin{cases} \alpha_k(i) & \text{if} \quad \begin{cases} \alpha_k(i) > 0, \\ z_{\alpha_k(i)} \in \{\vec{Z}\} \end{cases} \\ 0 & \text{if} \quad \text{otherwise} \end{cases}. \tag{23}$$

Using the notation in Eq. (9), if $\vec{Z}_k^{\alpha_k}$ is the vector of target-generated measurements in $\vec{Z}_k$ then

$$\{\vec{Z}^{\alpha}\} = \{\vec{Z}_k^{\alpha_k}\} \cap \{\vec{Z}\} \tag{24}$$

is the set of target-generated measurements in $\vec{Z}$; and $\{\vec{Z}\} - \{\vec{Z}^{\alpha}\}$ is the set of clutter measurements in $\vec{Z}$.

6.2 *MTA Likelihoods.* At time $t_k$ let: $p_{D,k}(x)$ be the sensor probability of detection; $f_k(z|x)$ the sensor measurement density; $\kappa_k(z)$ the intensity function of a Poisson clutter process; $\lambda_k = \int \kappa_k(z)dz$; $c_k(z) = \kappa_k(z)/\lambda_k$; and $\kappa_k^{\vec{Z}} = 1$ if $\vec{Z}$ is null and $\kappa_k^{\vec{Z}} = \prod_{z \in \{\vec{Z}\}} \kappa_k(z)$ otherwise. Let $\vec{Z} = (z_1, ..., z_m)$. Then $\vec{f}_k(\vec{Z}|\alpha^{\vec{Z}_k})$ is defined to be:

$$\vec{f}_k(z_1, ..., z_m|\alpha_k) = \frac{e^{-\lambda_k}}{m!} \kappa_k^{(z_1, ..., z_m)} \prod_{i:\alpha(i)>0} \frac{g_i^k(z_{\alpha(i)})}{c_k(z_{\alpha(i)})} \tag{25}$$

where $\alpha(i)$ was defined in Eq. (23) and where, for $s_i^{k|k-1}(x) = s_i^{k|k-1}(x|\alpha_{1:k-1}, \vec{Z}_{1:k-1})$ and $i = 1, ..., n$,

$$g_i^k(z) = \frac{\int p_{D,k}(x)\ f_k(z|x)\ s_i^{k|k-1}(x)dx}{\int p_{D,k}(y)\ s_i^{k|k-1}(y)dy} \tag{26}$$

is the probability (density) that $z$ is generated by the target $i$ (with target detectability taken into account);[10] whereas $c_k(z)$ is the probability (density) that $z$ is a clutter measurement. The unitless ratio $g_i^k(z)/c_k(z)$ quantifies the degree to which measurement $z$ is due to target $i$ rather than clutter. If $\{\vec{Z}_k^{\alpha_k}\} \cap \{\vec{Z}\} = \emptyset$ (i.e., if $\alpha^{\vec{Z}}$ has no target-generated measurements) then

$$\vec{f}_k(\vec{Z}|\alpha_k) = \frac{e^{-\lambda_k}\ \kappa_k^{\vec{Z}}}{|\vec{Z}|!} = \vec{\kappa}_k(\vec{Z}) \tag{27}$$

[8]GLMB distributions [13] are far more general and accurate, because they employ weighted sums of such products (given that MTAs are not state variables).

[9]Strictly speaking, $p_{1|0}(\alpha^{\vec{Z}})$ is now a probability density and not a probability and thus should be written as $\vec{f}_{1|0}(\alpha^{\vec{Z}})$. This technicality will be ignored for the sake of conceptual clarity.

[10]Note that if $s_i^{k|k-1}(x) = \delta_{x_i}(x)$ and $p_{D,k}(x_i) > 0$ then $g_i^k(z) = f_k(z|x_i)$; whereas if $p_{D,k}(x_i) = 0$ then $g_i^k(z) = 0$.

where $\vec{\kappa}_k(\vec{Z})$ denotes the distribution of the Poisson clutter process.

It is verified in Section 11 that $\int \vec{f}_k(\vec{Z}|\alpha_k)d\vec{Z} = 1$ for any $\alpha_k \in \mathbb{A}^{\vec{Z}_k}$ and any $\vec{Z}_k$.

## 7. Markov Transitions for MTAs

7.1 *Predicted MTAs for Markov Transitions.* Let $\alpha_{k-1} : \{1,...,n\} \to \{0,1,...,m_{k-1}\}$ be given, so that the target-generated measurement-vector at time $t_{k-1}$ is $\vec{Z}_{k-1}^{\alpha_{k-1}} = (z_{\alpha_{k-1}(i)} : \alpha_{k-1}(i) > 0)$. Assume that the single-target measurement density at time $t_k$ has the form

$$f_k(z|x_k) = f_{\mathbf{V}_k}(z - \eta_k(x_k)) \tag{28}$$

where $\eta_k(x_k) \in \mathbb{Z}_0$ is a measurement function and $\mathbf{V}_k \in \mathbb{Z}_0$ is zero-mean additive measurement noise.

Let $\hat{x}_i^{k|k-1} = \arg\sup_{x \in \mathbb{X}_0} s_i^{k|k-1}(x)$ for $i = 1,...,n$ be the estimated states of the predicted targets. Then the vector of all predicted target-generated measurements at time $t_k$ is

$$\vec{Z}_k^{\circ} = (\eta_k(\hat{x}_i^{k|k-1}) : i = 1,...,n). \tag{29}$$

According to $\alpha_{k-1}$, the only targets $i$ that generated measurements at time $t_{k-1}$ were those such that $\alpha_{k-1}(i) > 0$. Assume that these targets are measurement-generating at both times $t_k$ and $t_{k-1}$.[11] Then the actual vector of predicted target-generated measurements at time $t_k$ is

$$\vec{Z}_{k|k-1}^{\circ} = (\eta_k(\hat{x}_i^{k|k-1}) : \alpha_{k-1}(i) > 0). \tag{30}$$

Given this, for any $\vec{Z} \in \vec{\mathbb{Z}}_0$, define the predicted MTA $\alpha : \{1,...,n\} \to \{0,1,...,m\}$ to be

$$\alpha(i) = \alpha^{\vec{Z}}(i) = \begin{cases} \alpha_{k-1}(i) & \text{if} \quad \begin{cases} \alpha_{k-1}(i) > 0, \\ \eta_k(\hat{x}_i^{k|k-1}) \in \{\vec{Z}\} \end{cases} \\ 0 & \text{if} \quad \text{otherwise} \end{cases} . \tag{31}$$

The set of target-generated measurements in $\vec{Z}$ according to $\alpha$ is, therefore, $\{\vec{Z}^{\circ\alpha}\} = \{\vec{Z}_{k|k-1}\} \cap \{\vec{Z}\}$; whereas the set of clutter measurements in $\vec{Z}$ is $\{\vec{Z}\} - \{\vec{Z}^{\alpha}\}$.

7.2 *Markov Density for MTAs.* Given this, for $\vec{Z} = (z_1,...,z_m)$ the MTA Markov transition is defined to be:

$$p_{k|k-1}(\alpha^{\vec{Z}}|\alpha_{k-1}) = \frac{e^{-\lambda_k}\ \kappa_k^{\vec{Z}}\ \bar{N}_m^{-1}}{m!} \prod_{i:\alpha(i)>0} \frac{g_i^k(z_{\alpha(i)})}{c_k(z_{\alpha(i)})} \tag{32}$$

where $\alpha = \alpha^{\vec{Z}}$ was defined in Eq. (31); where $g_i^k(z)$ is as defined in Eq. (26); and where $\bar{N}_{|\vec{Z}|}$ was defined in Section 4.1. If $\{\vec{Z}_{k|k-1}^{\circ}\} \cap \{\vec{Z}\} = \emptyset$ ($\alpha^{\vec{Z}}$ has no target-generated measurements) then

$$p_{k|k-1}(\alpha^{\vec{Z}}|\alpha_{k-1}) = \vec{\kappa}_k(\vec{Z})\ \bar{N}_{|\vec{Z}|}^{-1}, \tag{33}$$

which does not depend on $\alpha_{k-1}$ at all. Since $\vec{Z}_k$ is yet to be collected, the factor $\bar{N}_{|\vec{Z}|}^{-1}$ acknowledges the fact that no $\alpha^{\vec{Z}} \in \mathbb{A}^{\vec{Z}}$ can be preferred on an a priori basis, insofar as the prediction $\alpha_{k-1} \to \alpha^{\vec{Z}}$ is concerned.

It is verified in Section 12 that, for any $\alpha_{k-1} \in \mathbb{A}^{\vec{Z}_{k-1}}$,

$$\int \sum_{\alpha^{\vec{Z}} \in \mathbb{A}^{\vec{Z}}} p_{k|k-1}(\alpha^{\vec{Z}}|\alpha_{k-1})d\vec{Z} = 1. \tag{34}$$

[11] This is an "inertial" MTA transition model, by analogy with inertial target-motion models. That is, measurement-generating targets at $t_{k-1}$ tend to be measurement-generating at $t_k$.

## 8. Physical Reality of MTAs

Beyond these purely mathematical issues, a more serious physical one must be addressed: *Is an MTA physically real?* Stated differently: Is it physically reasonable to claim that specific measurements are generated by specific targets?

This is doubtful. An MTA is a heuristic extrapolation to general multitarget scenarios of the following special case. Suppose that we have a sensor with no missed or false detections. Further suppose that we have $n$ targets with Dirac-delta spatial distributions $s_1^k(x) = \delta_{x_1}(x)$,..., $s_n^k(x) = \delta_{x_n}(x)$ that are well-separated with respect to the noise resolution of the sensor, as specified by $L_z(x) = f_k(z|x)$. In this case it is intuitively clear that there is a permutation $\pi_0$ on $1,...,n$ such that target $i$ generated $z_{\pi_0(i)}$. This is because there is only a small probability that $z_{\pi_0(i)}$ could have originated with any target other than $x_i$. That is, $\pi = \pi_0$ with

$$\begin{aligned} \vec{f}_k(\vec{Z}|\pi_0) &= g_1^k(z_{\pi_0(1)}) \cdots g_n^k(z_{\pi_0(n)}) && (35) \\ &= f_k(z_{\pi_0(1)}|x_1) \cdots f_k(z_{\pi_0(n)}|x_n), && (36) \end{aligned}$$

is supremal among $\vec{f}_k(\vec{Z}|\pi)$ over all permutations $\pi$ on $1,...,n$. (Here, $g_i^k(z)$ was defined in Eq. (26); and Eq. (36) follows from Footnote 10.)

Suppose, more generally, that the targets are very close together. Then it is statistically impossible to maintain that any particular measurement is generated by any particular target. Such a claim becomes even more difficult to maintain if the sensor has missed and false detections. But the improbable becomes truly impossible when one considers the following. First, point-detections are discretized mathematical idealizations that have been heuristically extracted from a continuous but noise-contaminated sensor signature. Second, the targets are assumed to be in an idealized "goldilocks zone": not too near (in which case a single target can generate multiple detections) and not too distant (in which case multiple targets can generate a single detection).

Consequently, to insist that there is a "Bayes-optimal" MTA is to impose a phenomenologically questionable structure upon the modeling of the physical system. Instead, the most that can legitimately be asserted is the following: If targets with state-set $X$ are present, then there is an RFS probability (density) $f_k(Z|X)$ that they will generate a measurement-set $Z$—see Eq. (39). From this it follows that MTAs are not physically valid state representations of multitarget systems.

Is this claim refuted by the fact that the RFS-like analysis in Sections 5-7 demonstrates that MHT has a Bayesian interpretation? No, because Bayesian reasoning can be ap-plied to even completely imaginary scenarios (e.g., tracking unicorns).

## 9. MTAs in MHT vs. MTAs in LRFS

MTAs also occur in LRFS likelihoods $f(Z|X)$. But they do not, as in MTA theory, arise from heuristic intuition. Rather, they arise from a *mere change of notation* as part of a mathematically rigorous LRFS derivation based on a statistically and phenomenologically rigorous LRFS multitarget measurement model. In particular, the MTAs in $f(Z|X)$ do not occur as isolated entities but, rather, simultaneously within the terms of a summation (see Eq. (39) and [4, Eq. 7.21]).

Specifically, let $X = \{x_1, ..., x_n\}$ and $Z = \{z_1, ..., z_m\}$ with $|X| = n$ and $|Z| = m$. Then the LRFS multitarget measurement model and multitarget calculus is used to derive the following formula:[12]

$$f(Z|X) = e^{-\lambda} \sum_{W_0 \uplus W_1 \uplus ... \uplus W_n = Z} \kappa^{Z-W_0} (1-p_D)^X \cdot \left( \prod_{i=1}^{n} \frac{f_i(W_i)}{1-p_D(x_i)} \right) \quad (37)$$

where $\kappa(Z) = e^{-\lambda}\lambda^{|Z|}c^Z$; where $(1-p_D)^X = \prod_{\mathbf{x}\in X}(1-p_D(x))$ if $X \neq \emptyset$ and $(1-p_D)^X = 1$ otherwise; where the summation is taken over all $W_1, ..., W_n \subseteq Z$ such that $|W_1|, ..., |W_n| \leq 1$;[13] and where

$$f_i(W) = \begin{cases} 1 & \text{if} \quad W = \emptyset \\ \frac{p_D(x_i)\cdot f(z|x_i)}{1-p_D(x_i)} & \text{if} \quad W = \{z\} \end{cases} . \quad (38)$$

Now, for a given choice of $W_1, ..., W_n$, define $\alpha : \{1, ..., n\} \to \{0, 1, ..., m\}$ implicitly by $\{z_{\alpha(i)}\} = W_i$ if $W_i \neq \emptyset$ and $\alpha(i) = 0$ otherwise. Then $\alpha$ is an MTA. Conversely, given $\alpha$, define $W_i = \{\mathbf{z}_{\alpha(i)}\}$ if $\alpha(i) > 0$ and $W_i = \emptyset$ if otherwise. Thus there is a one-to-one correspondence between MTAs $\alpha$ and lists $W_1, ..., W_n \subseteq Z$ such that $|W_1|, ..., |W_n| \leq 1$.

After this change from set-theoretic notation to MTA notation, Eq. (37) becomes [4, Eq. 7.21]

$$f(Z|X) = \kappa(Z)\,(1-p_D)^X \cdot \sum_{\alpha} \prod_{i:\alpha(i)>0} \frac{p_D(x_i)\ f(z_{\alpha(i)}|x_i)}{\kappa(z_{\alpha(i)})\ (1-p_D(x_i))}. \quad (39)$$

Note that the right side is invariant with respect to the orderings $X = \{x_1, ..., x_n\}$ and $Z = \{z_1, ..., z_m\}$ chosen for the elements of $Z$ and $X$.

## 10. Conclusion: MHT vs. LRFS

It should be evident at this point that LRFS is not a mathematically obfuscated replication of MHT. LRFS is mathematically, statistically, and physically rigorous, whereas MHT is physically heuristic. LRFS-based MTTs can near-optimally and simultaneously track over a million 3D targets in real time in significant clutter using off-the-shelf computing equipment—a direct consequence of the LRFS reformulation of MTT as a MTBRF on LFSs.

LRFS with SMMM also includes the Bernoulli filter [11], [7, Sec. 14.7], [9] and "dyadic filter" [5], which are non-approximate ECF special cases of the MTBRF when target number cannot exceed 1 or 2, respectively. Neither looks anything like MHT. In particular, the Bernoulli filter is equivalent to a probability hypothesis density (PHD) filter with a nonstandard multitarget motion model.

Moreover, while MHT is limited to SMMM, RFS/LRFS systematically addresses diverse sensor phenomenologies:

1. sensor signature ("raw") data, such as superpositional or pixelized-imaging sensors [4, Chpts. 19, 20]; and
2. "nonstandard" measurements (features, attributes, natural-language statements, inference rules), as part of a general theory of measurements based on random closed sets (RCSs) [4, Chpt. 22].

## 11. Appendix: MTA Likelihood Normality

By Eq. (7),

$$\int \vec{f}_k(\vec{Z}|\alpha_k)d\vec{Z} = \sum_{j\geq 0} \int f_k(z_1, ..., z_j|\alpha_k)dz_1 \cdots dz_j. \quad (40)$$

Abbreviate $\vec{W}_j = (z_1, ..., z_j)$. Let $\gamma_j : \{1, ..., n\} \to \{0, 1, ..., j\}$ be the extension of $\alpha_k$ to $\vec{W}_j$ as defined in Eqn. (23). Then from Eqs. (9, 24, 25) and $\kappa_k^{\vec{W}_j} = \lambda_k^j c_k^{\vec{W}_j}$ and $\{\vec{Z}^{\gamma_j}\} = \{\vec{Z}^{\alpha_k}\} \cap \{\vec{W}_j\}$ the desired result follows:

$$\int \vec{f}_k(\vec{Z}|\alpha_k)d\vec{Z} = e^{-\lambda_k} \sum_{j\geq 0} \frac{\lambda_k^j}{j!} \int \left( \prod_{l=1}^{j} c_k(z_l) \right) \cdot \left( \prod_{i:\gamma_j(i)>0} \frac{g_i^k(z_{\gamma_j(i)})}{c_k(z_{\gamma_j(i)})} \right) dz_1 \cdots dz_j \quad (41)$$

$$= e^{-\lambda_k} \sum_{j\geq 0} \frac{\lambda_k^j}{j!} \int \left( \prod_{l:z_l \in \{\vec{W}_j\} - \{\vec{W}_j^{\gamma_j}\}} c_k(z_l) \right) \cdot \left( \prod_{i:\gamma_j(i)>0} g_i^k(z_{\gamma_j(i)}) \right) dz_1 \cdots dz_j \quad (42)$$

$$= e^{-\lambda_k} \sum_{j\geq 0} \frac{\lambda_k^j}{j!} \left( \prod_{l:z_l \in \{\vec{W}_j\} - \{\vec{W}_j^{\gamma_j}\}} \int c_k(z_l)dz_l \right) \cdot \left( \prod_{i:\gamma_j(i)>0} \int g_i^k(z_{\gamma_j(i)})dz_{\gamma_j(i)} \right) \quad (43)$$

$$= e^{-\lambda_k} \sum_{j\geq 0} \frac{\lambda_k^j}{j!} = 1. \quad (44)$$

[12] For conceptual clarity the unlabeled RFS case is considered here. The LRFS case is similar but slightly more complicated.

[13] In which case $W_0 = Z - (W_1 \cup ... \cup W_n)$.

## 12. Appendix: MTA Markov Transition Normality

By Eq. (7),

$$\int \sum_{\alpha^{\vec{Z}} \in \mathbb{A}^{\vec{Z}}} p_{k|k-1}(\alpha^{\vec{Z}}|\alpha_{k-1}) d\vec{Z} \quad (45)$$

$$= \sum_{j \geq 0} \int \sum_{\alpha^{\vec{Z}} \in \mathbb{A}^{\vec{Z}}} p_{k|k-1}(\alpha^{(z_1,...,z_j)}|\alpha_{k-1}) dz_1 \cdots dz_j.$$

Abbreviate $\vec{W}_j = (z_1, ..., z_j)$. Let $\gamma_j : \{1, ..., n\} \to \{0, 1, ..., j\}$ be the predicted MTA as defined in Eq. (31). Then from Eqs. (30, 31) and $\{\vec{W}_j^{\gamma_j}\} = \{\vec{Z}^{\circ}_{k|k-1}\} \cap \{\vec{W}_j\}$ the desired result follows:

$$= \sum_{j \geq 0} \sum_{\gamma_j \in \mathbb{A}^{W_j}} \frac{e^{-\lambda_k} \lambda_k^j \bar{N}_j^{-1}}{j!} \int \left( \prod_{l=1}^{j} c_k(z_l) \right) \quad (46)$$

$$\cdot \left( \prod_{i:\gamma_j(i)>0} \frac{g_i^k(z_{\gamma_j(i)})}{c_k(z_{\gamma_j(i)})} \right) dz_1 \cdots dz_j$$

$$= \sum_{j \geq 0} \sum_{\gamma_j \in \mathbb{A}^{W_j}} \frac{e^{-\lambda_k} \lambda_k^j \bar{N}_j^{-1}}{j!} \int \left( \prod_{l:z_l \in \{\vec{W}_j\} - \{\vec{W}_j^{\gamma_j}\}} c_k(z_l) \right)$$

$$\cdot \left( \prod_{i:\gamma_j(i)>0} g_i^k(z_{\gamma_j(i)}) \right) dz_1 \cdots dz_j \quad (47)$$

$$= \sum_{j \geq 0} \sum_{\gamma_j \in \mathbb{A}^{W_j}} \frac{e^{-\lambda_k} \lambda_k^j \bar{N}_j^{-1}}{j!} \left( \prod_{l:z_l \in \{\vec{W}_j\} - \{\vec{W}_j^{\gamma_j}\}} \int c_k(z_l) dz_l \right)$$

$$\cdot \left( \prod_{i:\gamma_j(i)>0} \int g_i^k(z_{\gamma_j(i)}) dz_{\gamma_j(i)} \right) \quad (48)$$

$$= \sum_{j \geq 0} \frac{e^{-\lambda_k} \lambda_k^j \sum_{\gamma_j \in \mathbb{A}^{W_j}} \bar{N}_j^{-1}}{j!} = \sum_{j \geq 0} \frac{e^{-\lambda_k} \lambda_k^j}{j!} = 1. \quad (49)$$